\documentclass[aip, cp, amsmath, amssymb, reprint]{revtex4-2}

\usepackage[pdftex]{graphicx}
\usepackage{subfigure}
\usepackage{float}
\usepackage{indentfirst}
\usepackage{dcolumn}
\usepackage{bm}
\usepackage[utf8]{inputenc}
\usepackage[T1]{fontenc}
\usepackage{mathptmx}
\graphicspath{{./figures/}}
\usepackage{natbib}

\begin{document}

\title{Hairy Black Holes in Dilatonic Einstein-Gauss-Bonnet Theory}

\author{Bum-Hoon Lee}
\email{bhl@sogang.ac.kr}
\affiliation{Department of Physics\text{,} Sogang University\text{,} Seoul 04107\text{,} Korea}
\affiliation{Center for Quantum Spacetime\text{,} Sogang University\text{,} Seoul 04107\text{,} Korea}

\author{Hocheol Lee}
\email{insaying@sogang.ac.kr}
\affiliation{Department of Physics\text{,} Sogang University\text{,} Seoul 04107\text{,} Korea}
\affiliation{Center for Quantum Spacetime\text{,} Sogang University\text{,} Seoul 04107\text{,} Korea}

\author{Wonwoo Lee}
\email{warrior@sogang.ac.kr}
\affiliation{Center for Quantum Spacetime\text{,} Sogang University\text{,} Seoul 04107\text{,} Korea}

\begin{abstract}
We study black hole solutions in dilatonic Einstein-Gauss-Bonnet theory with a coupling constant $\alpha$ between the dilaton field and the Gauss-Bonnet term. In a previous study, we considered the black hole with the vanishing cosmological constant in this theory and constructed the hairy black hole solution with the negative $\alpha$. In this study, we present black hole solutions numerically with various physical properties in anti-de Sitter spacetime. We describe the procedure for constructing the black hole solutions in detail.
\end{abstract}

\maketitle

\section{Introduction \label{sec1}}
	In the very early universe approaching the Planck scale, the Einstein theory of gravitation could be incomplete. For this reason, we could think of the correction by higher-order curvature terms as one of the trials for this one. Among higher-order curvature terms, we consider the Gauss-Bonet (GB) term. That is a specific combination of square terms of the curvature. The GB term corresponds to the topological term that does not affect the dynamics in four-dimensional spacetime. If we consider a nonminimal coupling between the dilaton field and the GB term, it affects the equations of motion. Even with the introduction of this term, we still have second-order equations of motion \cite{Horndeski:1974wa}. The dilatonic-Einstein-Gauss-Bonnet theory (hereafter DEGB theory) is the simplest and the most natural extension of Einstein's one that it does not make a ghost state \cite{Ostrogradsky:1850fid, Woodard:2015zca, Nojiri:2018ouv}.

	Another characteristic of the DEGB theory is that a black hole solution has scalar hair outside the horizon \cite{Kanti:1995vq, Torii:1996yi}. The well-known black hole no-hair theorem \cite{Ruffini:1971bza} in Einstein-Scalar theory was established under several assumptions \cite{Bekenstein:1972ny, Bekenstein:1995un}. In the DEGB theory, the black hole no-hair theorem can be easily evaded since some assumptions are not satisfied or not applicable \cite{Kanti:1995vq, Herdeiro:2015waa, Antoniou:2017acq}. We have expanded the evasion of the black hole no-hair theorem with the negative $\alpha$ by constructing the new integral constraint equation, which allows the existence of the hairy black hole solution with the arbitrary sign of $\alpha$ \cite{Lee:2018zym}.

	The black hole no-hair theorem was conjectured in the Einstein-Maxwell theory with ordinary matters. Ordinary matters will eventually be absorbed into the black hole by gravitational interaction. However, black holes in nature coexist with matter fields represented by dark matter and dark energy \cite{Kiselev:2002dx, Cho:2017nhx, Kim:2019hfp}. It is well known that they are different from ordinary matters as we know them. For this reason, it is meaningful to introduce matter fields that do not belong to the standard particle model even the density is much less than that of a mass, find a solution that coexists with a black hole, and analyze their properties \cite{Kim:2004is, Kim:2007iv, Ko:2016dxa, Kang:2019uuj, Lee:2019ums}. As a result, it is worthwhile to study black holes with scalar hairs outside of a horizon as a simple model \cite{Guo:2008eq, Ohta:2009tb, Sotiriou:2013qea, Sotiriou:2014pfa, Doneva:2017bvd, Silva:2017uqg, Herdeiro:2018wub, Myung:2018iyq, Minamitsuji:2018xde, Bakopoulos:2018nui, Stetsko:2018fzt, Khodadi:2020jij, Doneva:2020qww, Kawai:2021edk}.

	We present black hole solutions numerically and study their properties in the DEGB theory. We construct black hole solutions in two cases: those without the cosmological constant and the negative cosmological constant. In this article, we focus on the procedure for finding the proper black hole solutions in anti-de Sitter (AdS) with the dilaton field decaying rapidly like an exponential function numerically and analyzing them.

	The paper is organized as follows: In the next section, we set up the basic framework. We describe the procedure of solving the equations of motion in detail. In third section, we present black hole solutions numerically and analyze their properties such as the energy-momentum tensor and energy conditions. We summarize and discuss our results in the last section.

\section{Hairy Black Hole Solutions in the DEGB Theory \label{sec2}}
	We numerically construct the asymptotically AdS black hole solution in DEGB theory. The numerical solutions and properties of those black holes are presented.
	
\subsection{The Model \label{sec2-1}}
	Let us consider the action with the GB term:
\begin{equation}
	S = \int_{\mathcal M} d^4 x \sqrt{-g} \left[\frac{R-2 p(\Phi) \Lambda}{2 \kappa} - \frac{1}{2}{\nabla^\mu} \Phi {\nabla_\mu} \Phi + q(\Phi) R^2_{\rm GB} \right] + S_b \,,
	\label{action}
\end{equation}
	where $g = \det g_{\mu\nu}$, $\kappa \equiv 8\pi G$, $p(\Phi) = e^{\lambda \Phi}$, $q(\Phi) = \alpha e^{\gamma \Phi}$ and $R$ denotes the scalar curvature of the spacetime $\mathcal M$\,. The negative cosmological constant, $\Lambda <0$, is considered for the AdS geometry. The GB term is given by $R^{2}_{\rm GB} = R^2 - 4 R_{\mu\nu} R^{\mu\nu} + R_{\mu\nu\rho\sigma} R^{\mu\nu\rho\sigma}$\,. The dilaton field $\Phi\,$ is coupled with the cosmological constant $\Lambda$ and the GB term $R^2_{\rm GB}$ by the coupling functions $p(\Phi)$ and $q(\Phi)$. The parameters $\alpha$, $\gamma$ and $\lambda$ are constants. The second term on the right-hand side $S_b$ is the boundary term \cite{Gibbons:1976ue, Myers:1987yn, Davis:2002gn}.
	
From the action Eq. \eqref{action}, the dilaton field equation and Einstein's equation are
\begin{eqnarray}
	0 &=& \nabla^2 \Phi + \dot{q}(\Phi) R_{GB}^2 - \frac{\dot{p}(\Phi) \Lambda}{\kappa}
	\\
	&=& \frac{1}{\sqrt{-g}} \partial_{\mu} [\sqrt{-g} g^{\mu \nu} \partial_{\nu} \Phi] + \dot{q}(\Phi) R_{GB}^2 - \frac{\dot{p}(\Phi) \Lambda}{\kappa} \,,
	\label{df}
	\\
	R_{\mu\nu} - \frac{1}{2} R g_{\mu\nu} &=& \kappa T_{\mu \nu}
	\\
	&=& \kappa \left( \partial_{\mu} \Phi \partial_{\nu} \Phi - \frac{1}{2} g_{\mu\nu} \partial_{\rho} \Phi \partial^{\rho} \Phi + T_{\mu\nu}^{\rm GB} - \frac{p(\Phi) \Lambda g_{\mu\nu}}{\kappa} \right) \label{EMT}
\end{eqnarray}
	where the dot notation denotes the derivative with respect to $\Phi$. The GB term contributes to the energy-momentum tensor as
\begin{eqnarray}
	T_{\mu \nu}^{\rm GB} &=& 4 R \nabla_\mu \nabla_\nu q(\Phi) + 8 R_{\mu \nu} \nabla^2 q(\Phi) - 8 R_{\mu \rho \nu \sigma} \nabla^\rho \nabla^\sigma q(\Phi) - 16 \nabla_\rho \nabla_{(\mu} q(\Phi) {R^\rho}_{\nu)} \nonumber
	\\
	&& + \left( 8 R^{\rho \sigma} \nabla_\rho \nabla_\sigma q(\Phi) - 4 R \nabla^2 q(\Phi) \right) g_{\mu \nu} \,.
	\label{TGB}
\end{eqnarray}

	Let us consider the spherically symmetric static metric as follows:
\begin{equation}
	ds^2 = - e^{X(r)} dt^2 + e^{Y(r)}  dr^2 + r^2 (d \theta^2 + \sin^2 \theta d \phi^2) \,,
	\label{metric}
\end{equation}
	where the metric functions $X$ and $Y$ depend only on $r$. By substituting the metric components, the GB term turns out to be
\begin{equation}
	R_{GB}^2 = \frac{2}{r^2} e^{-2 Y} \left[ \left( 1 - e^Y \right) \left( 2 X'' + X'^2 \right) + \left( 3 - e^Y \right) X' Y' \right]
	\label{RGB}
\end{equation}
	where the prime notation denotes the derivatives with respect to $r$. Using Eq. \eqref{EMT}, the components of the energy-momentum tensor $T^{\mu}_{\ \nu}$ are trun into
\begin{subequations}
\begin{eqnarray}
	T^t_{\ t} &=& \frac{8 \kappa \dot{q}}{r^2} e^{-2 Y} \left( 1 - e^Y \right) \Phi'' + \frac{\kappa}{2 r^2} e^{-2 Y} \left[ e^Y r^2 + 16 \ddot{q} \left( 1- e^Y \right) \right] \Phi'^2 \nonumber
	\\
	&& - \frac{4 \kappa \dot{q}}{r^2} e^{-2 Y} \left( 3 - e^Y \right) Y' \Phi' - p \Lambda \,,
	\label{Ttt}
	\\
	T^r_{\ r} &=& \frac{1}{2} \kappa e^{-Y} \Phi'^2 + \frac{4 \kappa \dot{q}}{r^2} e^{-2 Y} \left( 3 - e^Y \right) X' \Phi' - p \Lambda \,,
	\label{Trr}
	\\
	T^\theta_{\ \theta} &=&  T^\phi_{\ \phi} \nonumber
	\\
	&=& \frac{4 \kappa \dot{q}}{r} e^{-2 Y} X' \Phi'' + \frac{\kappa}{2 r} e^{-2 Y} \left( 8  \ddot{q} X' - r e^Y \right) \Phi'^2 + \frac{2 \kappa \dot{q}}{r} e^{-2 Y} \left( 2 X'' + X'^2 - 3 X' Y' \right) \Phi - p \Lambda \,.
	\label{Tthth}
\end{eqnarray}
\end{subequations}
    \newpage
    The dilaton field and Einstein's equations change into \cite{Kanti:1995vq} (with corrections of equations of motion in \cite{Khimphun:2016gsn} but without changing the main results)
\begin{subequations}
\begin{eqnarray}
	0 &=& \Phi'' + \frac{1}{2} \left( \frac{4}{r} + X' - Y' \right) \Phi' - \frac{2 \dot{q}}{r^2} e^{-Y} \left[ \left( 1 - e^Y \right) \left( 2 X'' + X'^2 \right) -  \left( 3 - e^Y \right) X' Y' \right] - \frac{\dot{p} \Lambda}{\kappa} e^{Y} \,,
	\label{eq:df}
	\\
	0 &=& \Phi'' + \left[ \frac{\ddot{q}}{\dot{q}} - \frac{r^2 e^Y}{16 \dot{q} \left( 1 - e^Y \right)} \right] \Phi'^2 - \frac{3 - e^Y}{2 \left( 1 - e^Y \right)} Y' \Phi' + \frac{e^Y \left[ r Y' + \left( 1-p \Lambda r^2 \right) e^Y -1 \right]}{8 \kappa \dot{q} \left( 1 - e^Y \right)} \,,
	\label{eq:tt}
	\\
	0 &=& \Phi'^2 + \frac{8 \dot{q}}{r^2} e^{-Y} \left( 3 - e^Y \right) X' \Phi' - \frac{2 \left[ r X' - \left( 1 - p \Lambda r^2 \right) e^Y + 1 \right]}{\kappa r^2} \,,
	\label{eq:rr}
	\\ 
	0 &=& \Phi'' + \left( \frac{\ddot{q}}{\dot{q}} - \frac{r e^Y}{8 \dot{q} X'} \right) \Phi'^2 + \left( \frac{X''}{X'} + \frac{X' - 3 Y'}{2} \right) \Phi' - \frac{e^Y \left[ r (2 X'' + 4 p \Lambda e^Y) - 2 Y' \right]}{16 \kappa \dot{q} X'} \nonumber
	\\
	&& + \frac{e^Y \left[ r (X' - Y') + 2 \right]}{16 \kappa \dot{q}} \,.
	\label{eq:qq}
\end{eqnarray}
\end{subequations}
	When $\Lambda$ is vanishing, equations of motion are reduced those with correcting a typo in Eq. \eqref{eq:rr} in \cite{Lee:2018zym} and changing with $\gamma \rightarrow -\gamma$ in \cite{Ahn:2014fwa, Ahn:2017rci}. Eq. \eqref{eq:rr} can be solved in terms of $Y$ as follows:
\begin{equation}
	e^Y = \frac{A \pm \sqrt{A^2 - B}}{4 \left(1 - p \Lambda r^2 \right)} \,,
	\label{eY}
\end{equation}
	where $A = -\kappa r^2 \Phi'^2 + 8 \kappa \dot{q} X' \Phi' + 2 r X' + 2$, $B = 192 \kappa \dot{q} \left(1 - p \Lambda r^2 \right) X' \Phi'$. We should take the positive sign to be valid at the near horizon, which is described below. Using Eq. \eqref{eY}, Eq. \eqref{eq:tt} and \eqref{eq:qq} reduce to the coupled second-order differential equations of $X''$ and $\Phi''$ as follows:
\begin{subequations}
\begin{eqnarray}
	X'' &=& \mathcal{X}(X, X', \Phi, \Phi') \,,
	\label{X''}
	\\
	\Phi'' &=& \mathcal{P}(X, X', \Phi, \Phi') \,.
	\label{Phi''}
\end{eqnarray}
\end{subequations}
	Those are the core equations that need to be solved by numerical calculation.

\subsection{Near Horizon Behavior} \label{Sec 2.2}
	For the numerical calculation of the coupled second-order differential equations, we have to determine the $4$ initial values, $X(r_h)$, $X'(r_h)$, $\Phi(r_h)$ and $\Phi'(r_h)$ where $r_h$ is the horizon radius. At the horizon, the metric component, e.g., $g_{tt}$, is zero. Therefore, the metric components and the dilaton field can be expanded in the near horizon limit by the length parameter $\delta r = r - r_h$ as follows:
\begin{subequations}
\begin{eqnarray}
	e^{X} &=& 0 + x_1 \delta r + \frac{1}{2} x_2 \delta r^2 + \mathcal{O}(\delta r^3) \,,
	\label{neX}
	\\
	e^{-Y} &=& 0 + y_1 \delta r + \frac{1}{2} y_2 \delta r^2 + \mathcal{O}(\delta r^3) \,,
	\\
	\Phi &=& \Phi_h + \Phi_h' \delta r + \frac{1}{2} \Phi_h'' \delta r^2 + \mathcal{O}(\delta r^3)
\end{eqnarray}
\end{subequations}
	where the subscript $h$ means the value at the horizon $r_h$. By differentiating Eq. \eqref{neX},
\begin{subequations}
\begin{eqnarray}
	X' e^X &=& x_1 + x_2 \delta r + \mathcal{O}(\delta r^2) \,, 	\\
    X' &=& \frac{1}{\delta r} + \mathcal{O}(1) \,. \label{nX'}
\end{eqnarray}
\end{subequations}
	The Eq. \eqref{nX'} shows that $X'$ diverges at the horizon. Due to the assumptions, we expand Eq. \eqref{eY} into terms of $X'$ as follows:
\begin{equation}
	e^Y = \frac{r + 4 \kappa \dot{q} \Phi'}{1 - p \Lambda r^2} X' - \frac{4 \kappa \dot{q} \left( \kappa r^2 \Phi'^2 - 6 p \Lambda r^2 + 4 \right) \Phi' + r \left( \kappa r^2 \Phi'^2 - 2 \right)}{2 \left( 1 - p \Lambda r^2 \right) \left( r + 4 \kappa \dot{q} \Phi' \right)} + \mathcal{O}\left( \frac{1}{X'} \right) \,.
	\label{eYh}
\end{equation}
	From Eq. \eqref{eYh}, $X'$ and $r+4 \kappa \dot{q} \Phi'$ are positive in the near horizon. Note that if we choose the negative sign form of Eq. \eqref{eY}, $e^Y = \mathcal{O}(1)$. That is not the proper behavior that we assumed in the near horizon. Finally, the expanded core equations in the near horizon are given by
\begin{subequations}
\begin{eqnarray}
	X'' &=& \frac{D}{C} X'^2 + \mathcal{O}(X') \,,
	\label{nX''}
	\\
	\Phi'' &=& \left( r + 4 \kappa \dot{q} \Phi' \right) \frac{E}{\kappa C} X' + \mathcal{O}(1)\,,
	\label{nPhi''}
\end{eqnarray}
\end{subequations}
\begin{eqnarray}
	\text{where} \;\; C &=& \left( 1 - p \Lambda r^2 \right) \left( r^4 + 4 \kappa \dot{q} \Phi' r^3 + 64 \kappa p \dot{q}^2 \Lambda r^2 - 128 \kappa^2 p \dot{q}^3 \Lambda \Phi' r - 96 \kappa \dot{q}^2 \right) \,, \nonumber
	\\
	D &=& p \Lambda r^6 + 8 \kappa p \dot{q} \Lambda \Phi' r^5 + \left( 16 \kappa^2 p \dot{q}^2 \Lambda \Phi'^2 + 48 \kappa p^2 \Lambda^2 + 4 \dot{p} \dot{q} \Lambda - 1 \right) r^4 \nonumber
	\\
	&& - 8 \kappa \dot{q} \left( 1 - 4 \dot{p} \dot{q} \Lambda \right) \Phi' r^3 + 16 \kappa \dot{q}^2 \left[ \kappa \left( 1 - 4 \dot{p} \dot{q} \Lambda \right) \Phi'^2 + 8 p \Lambda \right] r^2 \nonumber
	\\
	&& - 256 \kappa^2 p \dot{q}^3 \Phi' r - 512 \kappa^3 p \dot{q}^4 \Lambda \Phi'^2 \,, \nonumber
	\\
	E &=& \kappa p \Lambda \Phi' r^5 + \left[ 4 \kappa p \dot{q} \left( \kappa \Phi'^2 - p \Lambda \right) + \dot{p} \right] \Lambda r^4 + \kappa \left( 32 \kappa p^2 \dot{q}^2 \Lambda^2 + 8 \dot{p} \dot{q} \Lambda - 1 \right) \Phi' r^3 \nonumber
	\\
	&& - 4 \kappa \dot{q} \left[ \kappa \left( 1 - 4 \dot{p} \dot{q} \Lambda \right) \Phi'^2 - 2 p \Lambda \right] r^2 - 96 \kappa^2 p \dot{q}^2 \Lambda \Phi' r - 4 \kappa \dot{q} \left( 32 \kappa^2 p \dot{q}^2 \Lambda \Phi'^2 + 3 \right) \,. \nonumber
\end{eqnarray}
	In Eq. \eqref{nPhi''}, $\Phi''$ diverges at the horizon because of $X'$. Therefore, only the numerator E should be zero to avoid the divergence of $\Phi''$. The value of $\Phi'$ at the horizon $r_h$ is derived from aforementioned condition as follows:
\begin{equation}
	\Phi_h' = -\frac{G \pm \left( 1 - p \Lambda r_h^2 \right) \sqrt H}{F} \,,
	\label{Phi'}
\end{equation}
\begin{eqnarray}
	\text{where} \;\; F &=& -8 \kappa \dot{q}_h \left[ p_h \Lambda r_h^4 - \left( 1 - 4 \dot{p}_h \dot{q}_h \Lambda \right) r_h^2 - 32 \kappa p_h \dot{q}_h^2 \Lambda \right] \,, \nonumber
	\\
	G &=& -p_h \Lambda r_h^5 - \left( 32 \kappa p_h^2 \dot{q}_h^2 \Lambda^2 + 8 \dot{p}_h \dot{q}_h \Lambda -1 \right) r_h^3 + 96 \kappa p_h \Lambda r_h \,, \nonumber
	\\
	H &=& r_h^6 + 128 \kappa p_h \dot{q}_h^2 \Lambda r_h^4 + 64 \kappa \dot{q}_h^2 \left( 16 \kappa p_h^2 \dot{q}_h^2 \Lambda^2 +12 \dot{p}_h \dot{q}_h \Lambda -3 \right) r_h^2 - 6144 \kappa^2 p_h \dot{q}_h \Lambda \,. \nonumber
	\label{H}
\end{eqnarray}
	We should choose the negative sign form of $\Phi_h'$ for $e^Y$ to be valid in the near horizon, $e^Y \approx X'$. To obtain the real value, $H$ must be a non-negative value, but one caveat is that the case where $H$ is zero should be avoided. If $H = 0$, $\Phi_h' = -\frac{G}{F}$ and then substitute $\Phi_h' = -\frac{G}{F}$ into \eqref{nPhi''}, $\Phi''$ reduce as follows:
\begin{equation}
	\Phi'' = \frac{r + 4 \kappa \dot{q} \Phi'}{8 \kappa \dot{q}} X' + \mathcal{O}(1) \,. \nonumber
\end{equation}
	Due to the reduction of fraction, still $X'$ remains and $\Phi''$ diverges even after the substitution. That is not proper behavior of $\Phi''$ in the near horizon. Thus, $H$ should be a positive value. The constraint of $4 + \kappa \dot{q} \Phi'$ and $H$, which are both have to be postive, lead the allowed region of $\Phi(r_h)$ with given parameters and horizon radius $r_h$. The substitution of $\Phi'(r_h)$ into Eqs. \eqref{nX''} and \eqref{nPhi''} reduces $X''$ and $\Phi''$ in the near horizon as follows:
\begin{subequations}
\begin{eqnarray}
	X'' &\approx& -X'^2 \,,
	\label{nX''2}
	\\
	\Phi'' &\approx& 0 \,.
	\label{nPhi''2}
\end{eqnarray}
\end{subequations}
	Then, $X'$ is derived from Eq. \eqref{nX''2},
\begin{equation}
	X'(r) \approx \frac{1}{\delta r}  \,.
	\label{nX'2}
\end{equation}
	This result recovers Eq. \eqref{nX'}, which is the original assumption in the near horizon. Using Eq. \eqref{nX''2} and \eqref{nX'2}, the components of the energy-momentum tensor also can be expanded in terms of $\delta r$ as follows:
\begin{subequations}
\begin{eqnarray}
	T^t_{\ t} = T^r_{\ r} &=& -\frac{p \Lambda r^3 + 4 \kappa \dot{q} \Phi'}{r^2 (r + 4 \kappa \dot{q} \Phi')} + \mathcal{O}(\delta r) \,,
	\label{TtthTrrh}
	\\
	T^\theta_{\ \theta} = T^\phi_{\ \phi} &=& \frac{4 \kappa \dot{q} (1 - p \Lambda r^2)^2 \Phi'}{2 r (r + 4 \kappa \dot{q} \Phi')^2} - p \Lambda + \mathcal{O}(\delta r)  \,.
	\label{Tththh}
\end{eqnarray}
\end{subequations}
	In the same way, the GB term $R^2_{\rm GB}$ in the near horizon is obtained by
\begin{equation}
	R^2_{GB} \approx \frac{4}{r^2} e^{-2 Y} X'^2 \,.
\end{equation}

\subsection{Asymptotic Analysis} \label{Sec 2.3}
	We expect the dilaton field decays rapidly where the radius is large from the horizon. Thus, we can approximate the metric to be the form of the Schwarzschild-AdS-like solution:
\begin{subequations}
\begin{eqnarray}
	e^X &=& 1 - \frac{2M}{r} - \frac{p(\Phi) \Lambda}{3} r^2 \,,
	\label{aeX}
	\\
	e^{-Y} &=& 1 - \frac{2M}{r} - \frac{p(\Phi) \Lambda}{3} r^2 \,.
	\label{aeY}
\end{eqnarray}
\end{subequations}
	Substitute Eqs. \eqref{aeX} and \eqref{aeY} into Eq. \eqref{eq:df}. Then, Eq. \eqref{eq:df} turns into
\begin{eqnarray}
	 0 &=& \left[ p \Lambda \left( 1 - \frac{4}{3} \dot{p} \dot{q} \Lambda \right) r^4 - 3 r^2 + 6 M \left( 1 - \frac{4}{3} \dot{p} \dot{q} \Lambda \right) r \right] \Phi'' \nonumber
	\\
	&& + \left\{ \dot{p} \Lambda \left[ 1 - \frac{4}{3} \left( \dot{p} \dot{q} + \frac{p \ddot{p} \dot{q}}{\dot{p}} \right) \Lambda^2 \right] - 8 M \ddot{p} \dot{q} \Lambda r \right\} \Phi'^2 \nonumber
	\\
	&& + \left[ 4 p \Lambda \left( 1 - \frac{8}{3} \dot{p} \dot{q} \Lambda \right) r^3 - 6 r + 6 M \left( 1 - \frac{8}{3} \dot{p} \dot{q} \Lambda \right) \right] \Phi' \nonumber
	\\
	&& + \frac{3 \dot{p} \Lambda}{\kappa} \left( 1 - \frac{8}{3} \frac{\kappa p^2 \dot{q}}{\dot{p}} \Lambda \right) r^2 - \frac{144 M^2 \dot{q}}{r^4} \,.
	\label{eq:adf}
\end{eqnarray}
	We consider only the highest order of $r$ and demand that $\Phi$ converges to constant value in the asymptotic region as power series: $\Phi = \Phi_\infty + \frac{\Phi_1}{r} + \frac{\Phi_2}{r^2} + \dotsm$. Then, Eq. \eqref{eq:adf} reduces as follows:
\begin{equation}
	0 = \left( 1 - \frac{8 \kappa p_\infty^2 \dot{q}_\infty}{3 \dot{p}_\infty} \Lambda \right) r^2 + \mathcal{O}(r)
	\label{eq:aLambda}
\end{equation}
	where subscript $\infty$ means the value in the asymptotic region, e.g., $p_\infty = p(\Phi_\infty)$. We can obtain the relation that should be satisfied among the parameters from Eq. \eqref{eq:aLambda}:
\begin{equation}
	\Lambda = \frac{3 \dot{p}_\infty}{8 \kappa p_\infty^2 \dot{q}_\infty} = \frac{3 \lambda}{8 \kappa \alpha \gamma} e^{-(\lambda + \gamma) \Phi_\infty} \,.
	\label{aLambda}
\end{equation}
    Using the above assumptions, the components of energy-momentum tensor in the asymptotic region are given by
\begin{subequations}
\begin{eqnarray}
	T^t_{\ t} &=& -p_\infty \Lambda \,,
	\label{aTtt}
	\\
	T^r_{\ r} &=& -p_\infty \Lambda \,,
	\label{aTrr}
	\\
	T^\theta_{\ \theta} &=& T^\phi_{\ \phi} \nonumber
	\\
	&=& -p_\infty \Lambda \,.
	\label{aTthth}
\end{eqnarray}
\end{subequations}
	In like manner, the GB term $R^2_{\rm GB}$ in the asymptotic region is obteined by
\begin{equation}
	R^2_{\rm GB} = \frac{8}{3} p_\infty^2\Lambda^2 \,.
	\label{aRGB}
\end{equation}

\subsection{Shifting and Rescaling} \label{Sec 2.4}
	To adjust the value of the dilaton field in the asymptotic region as $0$, we shift the dilaton field $\Phi(r) \rightarrow \tilde{\Phi}(r) = \Phi(r) - \Phi_\infty$. For the invariance of equations, the radius $r$ and cosmological constant $\Lambda$ must be rescaled to $\tilde{r}$ and $\tilde{\Lambda}$ with specific form by the coupling functions $p(\Phi) = e^{\lambda \Phi}$ and $q(\Phi) = \alpha e^{\gamma \Phi}$:
\begin{equation}
	r \rightarrow \tilde{r} = e^{-\frac{\gamma}{2} \Phi_\infty} r \,, \qquad \Lambda \rightarrow \tilde{\Lambda} = e^{(\lambda + \gamma) \Phi_\infty} \Lambda \,. \nonumber
\end{equation}
    The relations between the first derivative and second derivative of the arbitrary function $f$ with respect to $r$ and $\tilde{r}$ are given by
\begin{eqnarray}
	f'(r) &=& \frac{d}{dr} f(r(\tilde{r})) = \frac{d\tilde{r}}{dr} \frac{d}{d\tilde{r}} f(\tilde{r}) \equiv e^{-\frac{\gamma}{2} \Phi_\infty} f'(\tilde{r}) \,, \nonumber
	\\
	f''(r) &=& \frac{d}{dr} f'(r(\tilde{r})) = \frac{d}{dr} \left( e^{-\gamma \Phi_\infty / 2} f'(\tilde{r}) \right) = e^{-\frac{\gamma}{2} \Phi_\infty} \frac{d}{d\tilde{r}} \left( e^{-\frac{\gamma}{2} \Phi_\infty} f'(\tilde{r}) \right) \equiv e^{-\gamma \Phi_\infty} f''(\tilde{r}) \,. \nonumber
\end{eqnarray}
    From now on, the prime notation is redefined as derivative with respect to variable of the function. After the shifting and rescaling, the original functions are changed into the rescaled function as follows:
\begin{eqnarray}
	\Phi(r) &\rightarrow& \tilde{\Phi}(\tilde{r}) + \Phi_\infty \,, \qquad\quad\: X(r) \; \rightarrow \; X(\tilde{r}) \,, \qquad\qquad\qquad\! Y(r) \; \rightarrow \; Y(\tilde{r}) \,, \nonumber
	\\
	\Phi'(r) &\rightarrow& e^{-\frac{\gamma}{2} \Phi_\infty} \tilde{\Phi}'(\tilde{r}) \qquad\quad X'(r) \; \rightarrow \; e^{-\frac{\gamma}{2} \Phi_\infty} X'(\tilde{r}) \:\qquad\quad Y'(r) \; \rightarrow \; e^{-\frac{\gamma}{2} \Phi_\infty} Y'(\tilde{r}) \,, \nonumber
	\label{rescaling}
	\\
	\Phi''(r) &\rightarrow& e^{-\gamma \Phi_\infty} \tilde{\Phi}''(\tilde{r}) \,, \qquad\,\, X''(r) \; \rightarrow \; e^{-\gamma \Phi_\infty} X''(\tilde{r}) \,, \qquad\,\,\,\, Y''(r) \; \rightarrow \; e^{-\gamma \Phi_\infty} Y''(\tilde{r}) \,. \nonumber
\end{eqnarray}
	Then, the rescaled equations of motion are given by
\begin{eqnarray}
	0 &=& e^{-\gamma \Phi_\infty} \times \rm{Eq}. \; \eqref{eq:df} \,, \nonumber
	\\
	0 &=& e^{-\gamma \Phi_\infty} \times \rm{Eq}. \; \eqref{eq:tt} \,, \nonumber
	\\
	0 &=& e^{-\gamma \Phi_\infty} \times \rm{Eq}. \; \eqref{eq:rr} \,, \nonumber
	\\
	0 &=& e^{-\gamma \Phi_\infty} \times \rm{Eq}. \; \eqref{eq:qq} \,. \nonumber
\end{eqnarray}
	The rescaled equations of motion are only constant multiple of original equations of motion. Therefore, the equations of motion are invariant under the shift of the dilaton field $\Phi \rightarrow \tilde{\Phi} = \Phi - \Phi_\infty$ with the rescaling of $r \rightarrow \tilde{r} = e^{-\frac{\gamma}{2} \Phi_\infty} r$ and $\Lambda \rightarrow \tilde{\Lambda} = e^{(\lambda + \gamma) \Phi_\infty} \Lambda$. Since the shifting and rescaling are a set, after this section, the function with tilde symbol after this section is naturally the function with respect to $\tilde{r}$, e.g., $\tilde{X} = X(\tilde{r})$ and only if the dilaton field $\Phi$ with tilde symbol contains the shifting and the rescaling both, e.g., $\tilde{\Phi} = \tilde{\Phi}(\tilde{r}) - \Phi_\infty$.

\section{Numerical Solutions \label{sec3}}
    In this section, we present numerical solutions of black holes with the scalar hair. We take $r_h = 1$, $\kappa = 1$, $\gamma = -1$, $\alpha = -1$ without the cosmological constant, and $\alpha = 1$ with $\lambda = -\gamma$ and $\Lambda = \frac{3 \lambda}{8 \kappa \alpha \gamma}$, for simplicity.

\subsection{Black Hole in an Asymptotic Flat Spacetime \label{sec3-1}}
    We present numerical solutions of hairy black hole by solving Eqs. \eqref{X''} and \eqref{Phi''} in DEGB theory.
\begin{figure}[H]
	\centering
	\subfigure[][$-\tilde{g}_{tt}$ and $\tilde{g}_{rr}$ vs. $\tilde{r}$]
{\includegraphics[width = 0.4 \textwidth]{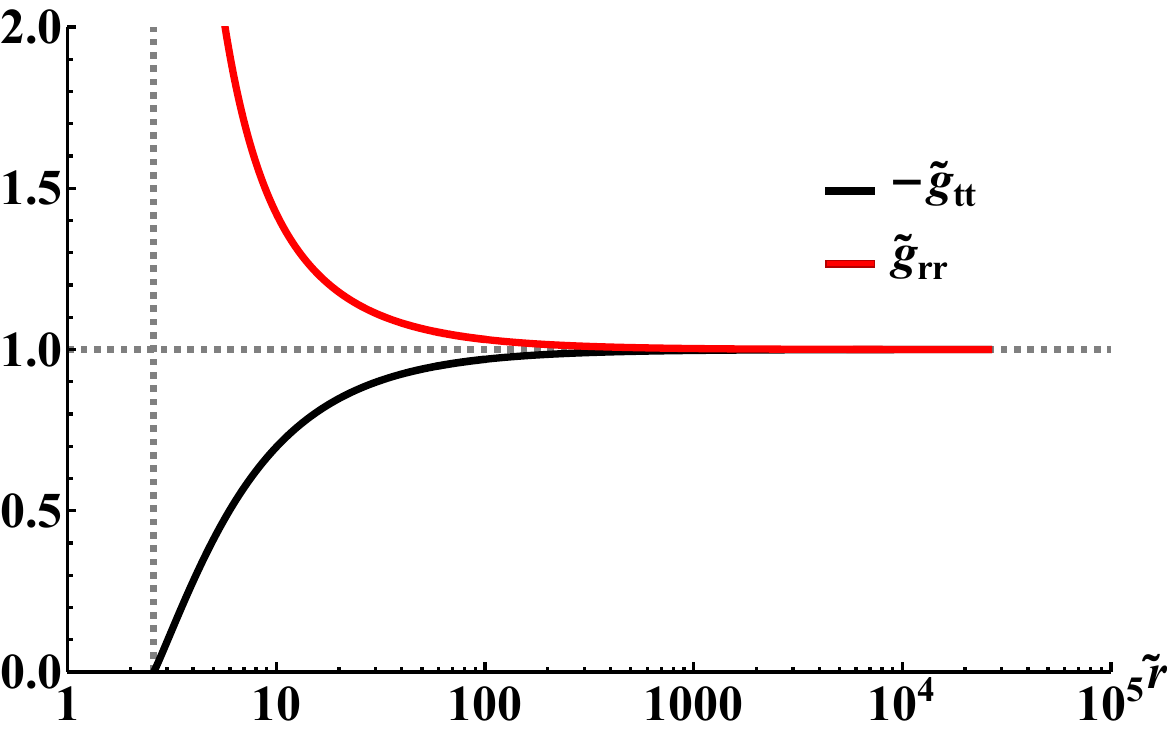}}
	\subfigure[][$\tilde{\Phi}$ vs. $\tilde{r}$]
{\includegraphics[width = 0.4 \textwidth]{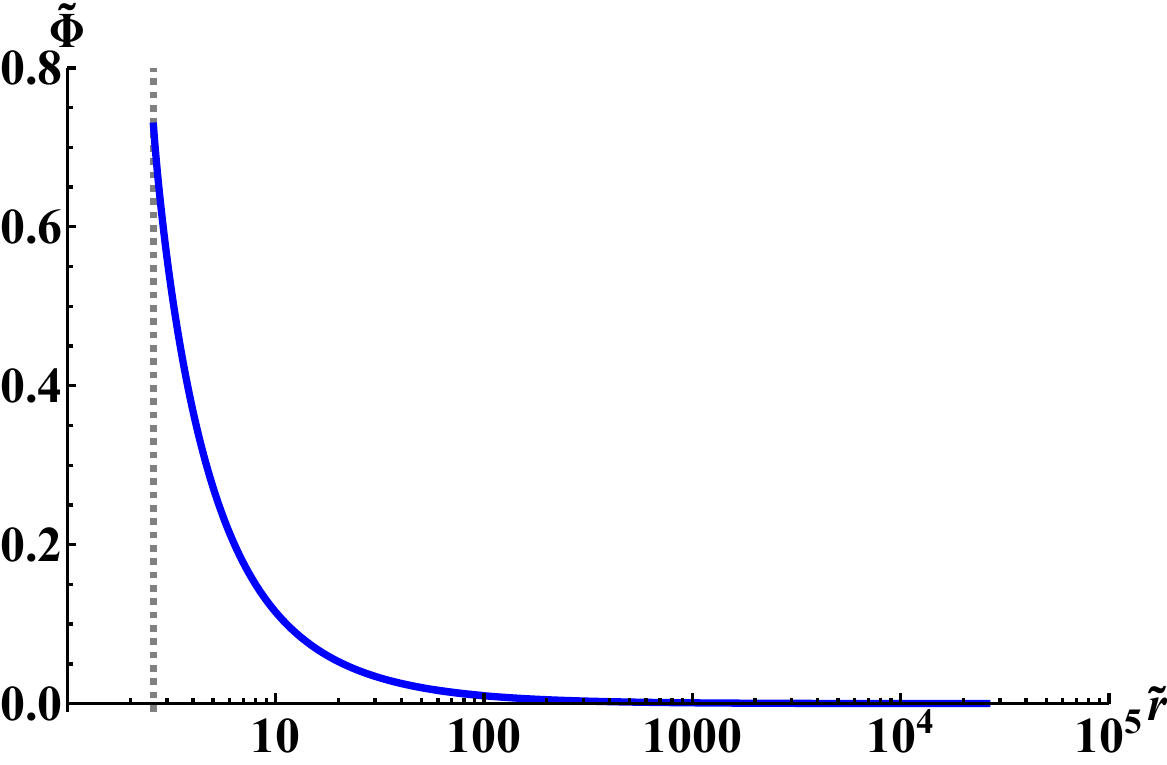}}
	\caption{The numerical solutions for metric components and the profile of the dilaton field.}
	\label{fig:EdSMetric_Phi}
\end{figure}
	Figure \ref{fig:EdSMetric_Phi} illustrates the metric components and the dilaton field for the black hole solution with respect to $\tilde{r}$. In Fig. \ref{fig:EdSMetric_Phi}(a), the black line indicates $-\tilde{g}_{tt}$ and the red line indicates $\tilde{g}_{rr}$. Both the black and red lines rapidly converge to $1$ as $\tilde{r}$ increases. In Fig. \ref{fig:EdSMetric_Phi}(b), the blue line indicates the dilaton field. The blue line rapidly converges to $0$ as $\tilde{r}$ increases. The vertical dashed gray line indicates the rescaled horizon radius $\tilde{r}_h$.
\begin{figure}[H]
	\centering
	\subfigure[][$\tilde{R}^2_{GB}$ vs. $\tilde{r}$]
{\includegraphics[width = 0.4 \textwidth]{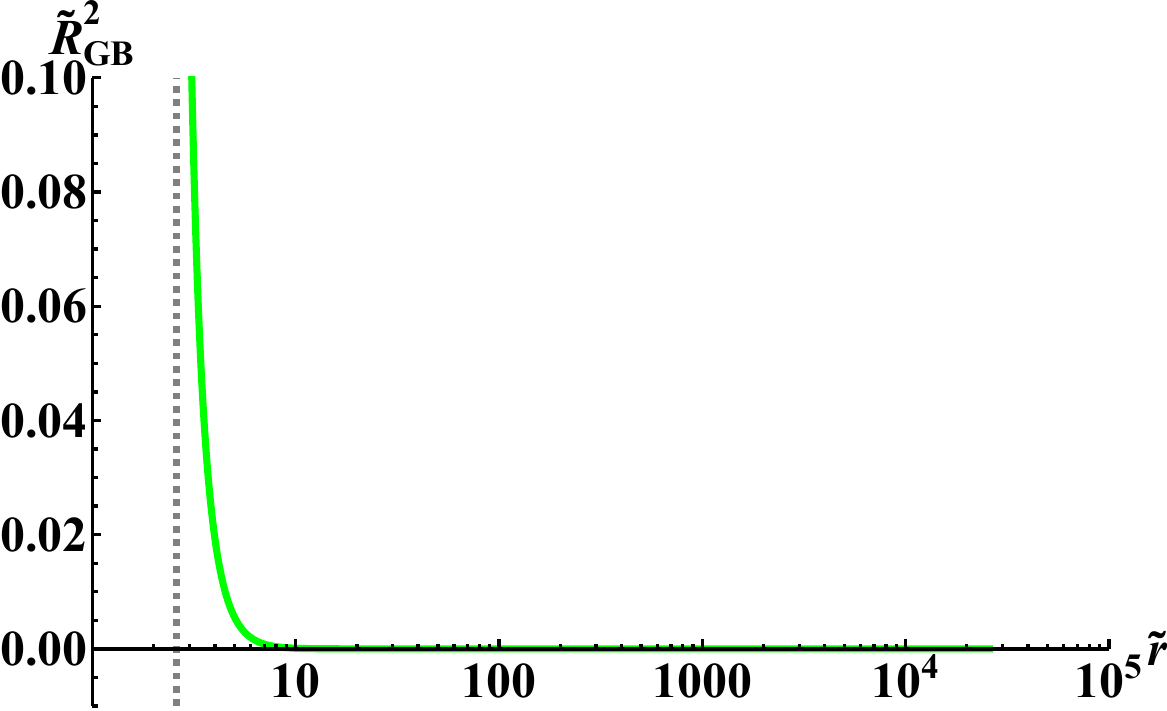}}
	\subfigure[][$2 \tilde{M}$ vs. $\tilde{r}$]
{\includegraphics[width = 0.4 \textwidth]{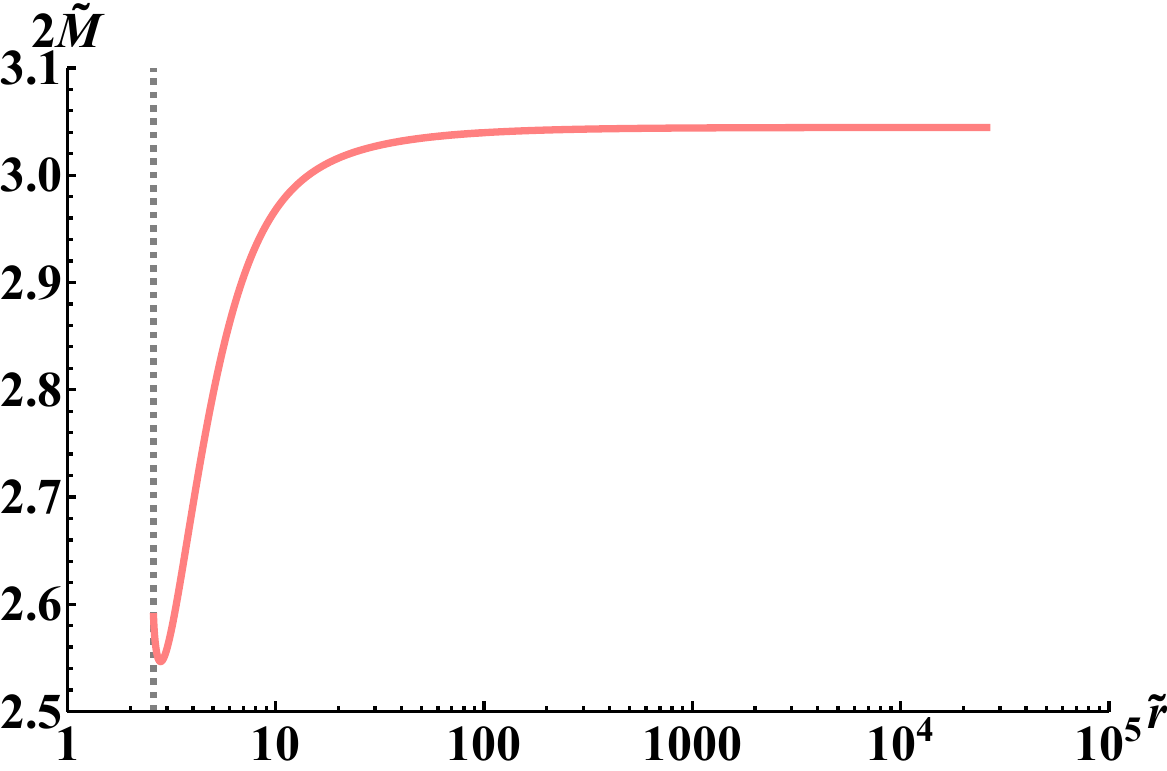}}
	\caption{The GB term $\tilde{R}^2_{GB}$ and mass function $2 \tilde{M}$.}
	\label{fig:EdSRGB_2M}
\end{figure}
	Figure \ref{fig:EdSRGB_2M} represents the GB term $\tilde{R}^2_{GB}$ and mass function $2 \tilde{M}$ with respect to $\tilde{r}$. In Fig. \ref{fig:EdSRGB_2M}(a), the GB term has a positive value in all regions of $\tilde{r}$ and has finite value at $\tilde{r}_h$. In Fig. \ref{fig:RGB_2M}(b), the mass function increases and then converges to some constant value. The effect of the decaying dilaton field fades out as $\tilde{r}$ increases.
\begin{figure}[H]
	\centering
	\subfigure[][$\tilde{T}^t_{\ t}, \, \tilde{T}^r_{\ r}, \, \tilde{T}^\theta_{\ \theta} \enspace vs. \enspace \tilde{r}$]
{\includegraphics[width = 0.4 \textwidth]{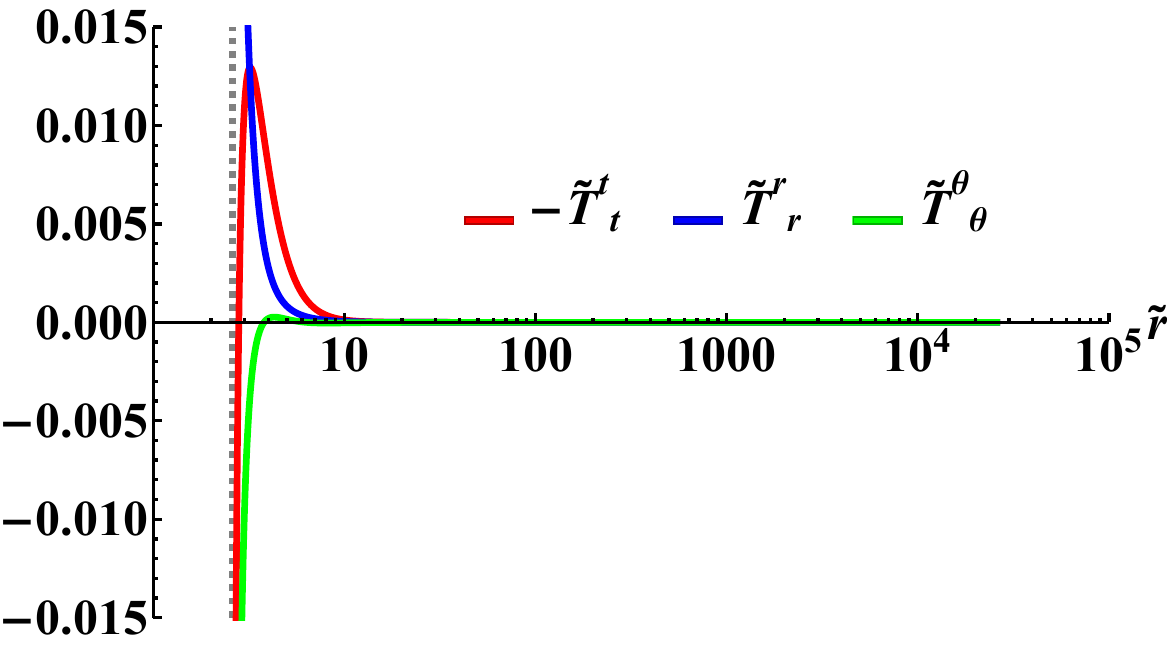}}
	\subfigure[][$-\tilde{T}^t_{\ t} + \tilde{T}^r_{\ r} \enspace vs.enspace \tilde{r}$]
{\includegraphics[width = 0.4 \textwidth]{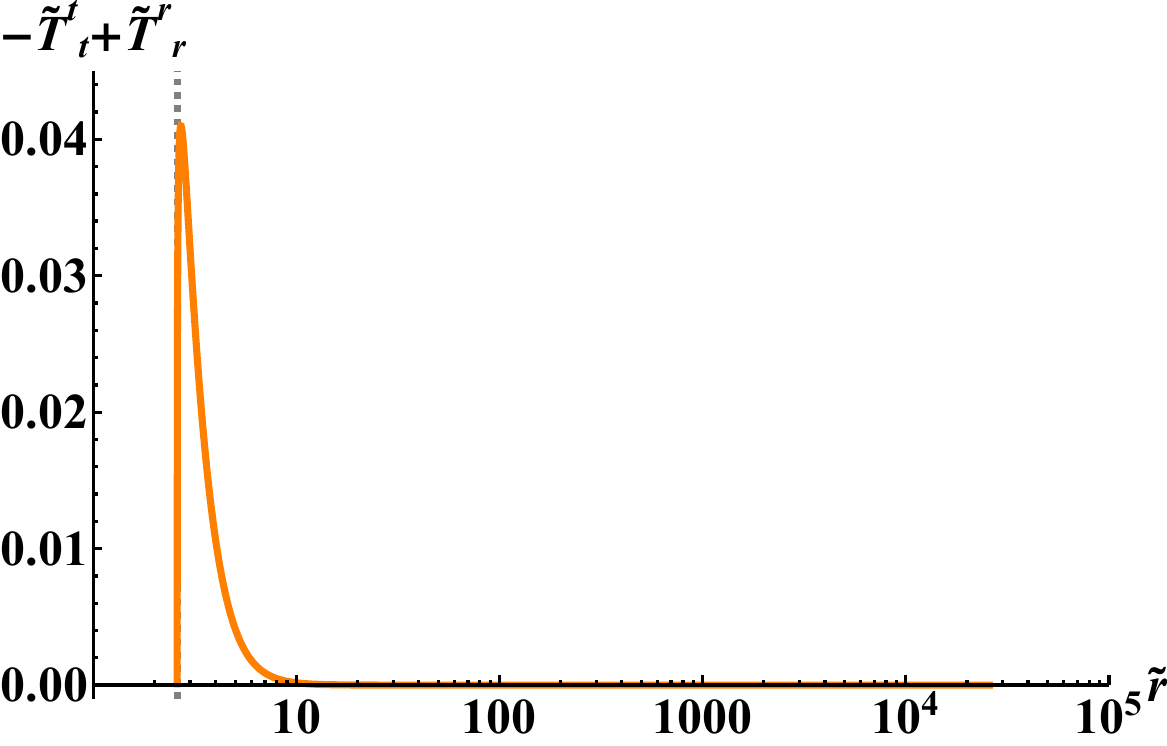}}
	\caption{The energy momentum tensor components and null energy condition.}
	\label{fig:EdSEMT_Null}
\end{figure}
	Figure \ref{fig:EdSEMT_Null} shows the components of the energy-momentum tensor and the checked value of whether the null energy condition is satisfied. In Fig \ref{fig:EdSEMT_Null}(a), the red line depicts the $(tt)$ component $- \tilde{T}^t_{\ t}$, the blue line depicts the $(rr)$ component $\tilde{T}^r_{\ r}$ and the green lines depict the ($\theta \theta$) component $\tilde{T}^\theta_{\ \theta}$. The negative part of $-\tilde{T}^t_{\ t}$ violates the key assumption of Bekenstein's novel no-hair theorem. In Fig. \ref{fig:EdSEMT_Null}(b), the orange line depicts $-\tilde{T}^t_{\ t} + \tilde{T}^r_{\ r}$ as known as null energy condition. All values are non-negative, which means the black hole solution satisfies the null energy condition in all regions.

\newpage
\subsection{Black Hole in an AdS Spacetime \label{sec3-2}}
\begin{figure}[H]
	\centering
	\subfigure[][$-\left( \tilde{k} \tilde{r}^2 \right)^{-1} \tilde{g}_{tt}$ and $\tilde{k} \tilde{r}^2 \tilde{g}_{rr}$ vs. $\tilde{r}$]
{\includegraphics[width = 0.43 \textwidth]{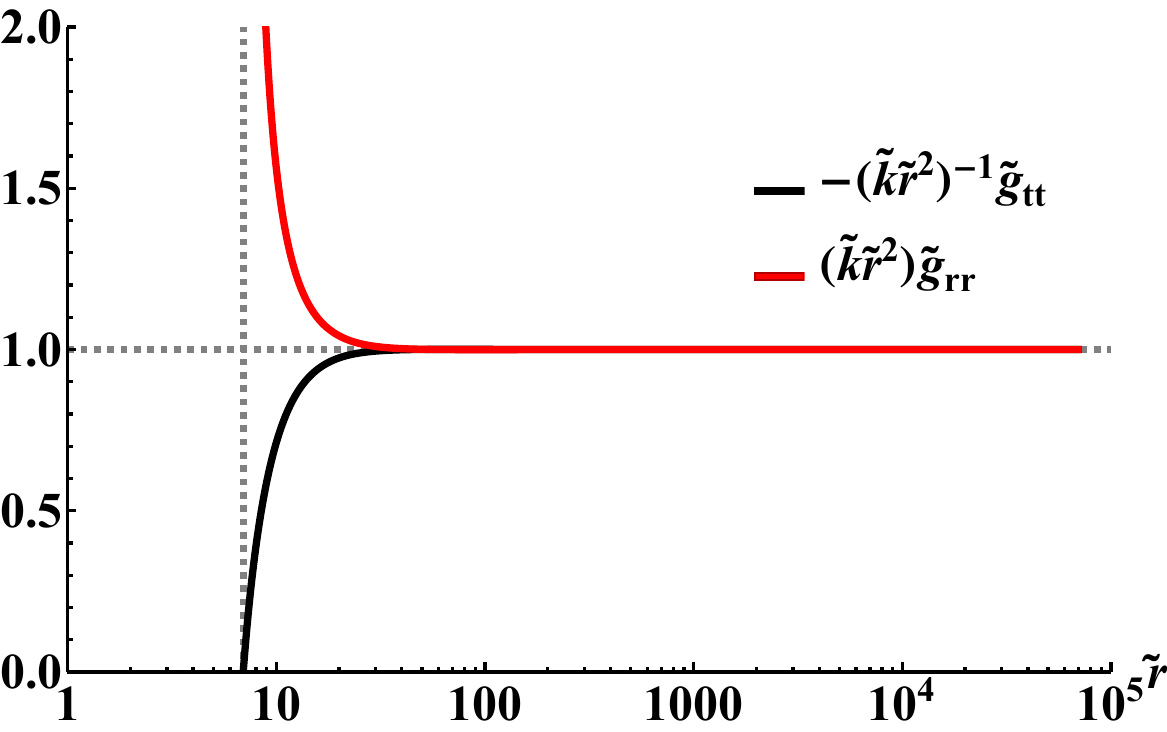}}
	\subfigure[][$\tilde{\Phi}$ vs. $\tilde{r}$]
{\includegraphics[width = 0.43 \textwidth]{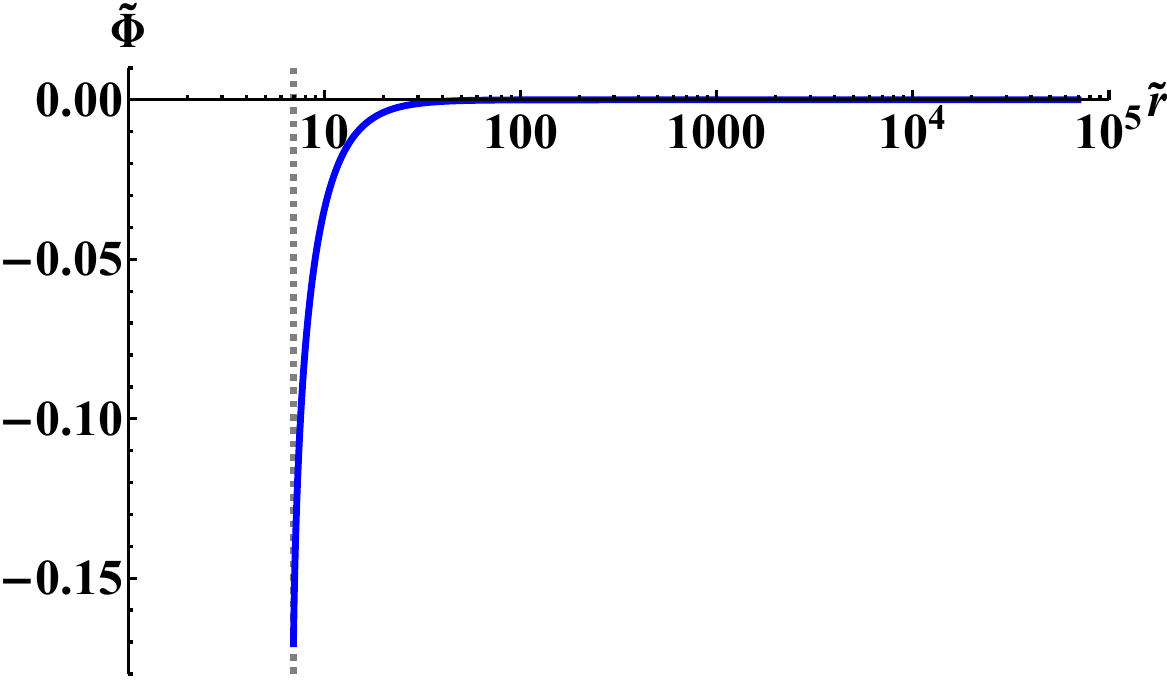}}
	\caption{The metric components and dilaton field for AdS black hole solution where $\tilde{k} = -\frac{\tilde{p} \tilde{\Lambda}}{3}$.}
	\label{fig:Metric_Phi}
\end{figure}
    Figure \ref{fig:Metric_Phi} illustrates the metric components and the dilaton field for AdS black hole solution with respect to $\tilde{r}$. In Fig. \ref{fig:Metric_Phi}(a), the black line indicates $-\left( \tilde{k} \tilde{r}^2 \right)^{-1} \tilde{g}_{tt}$ and the red line indicates $\tilde{k} \tilde{r}^2 \tilde{g}_{rr}$. Both the black and red lines rapidly converge to $1$ as $\tilde{r}$ increases. In Fig. \ref{fig:Metric_Phi}(b), the blue line indicates the dilaton field. The blue line rapidly converges to $0$ as $\tilde{r}$ increases. Therefore, we can conclude that the metric components and dilaton field fit the boundary conditions at the large distance from the horizon:
\begin{subequations}
\begin{eqnarray}
	-\left( \tilde{k}_\infty \tilde{r}_\infty^2 \right)^{-1} \tilde{g}_{tt}(\tilde{r}_\infty) &=& \left( -\frac{\tilde{p}_{\infty} \tilde{\Lambda}}{3} \tilde{r}_\infty^2 \right)^{-1} e^{\tilde{X}_\infty} \; \approx \; 1 \,, \nonumber
	\\
	\left( \tilde{k}_\infty \tilde{r}_\infty^2 \right) g_{rr}(\tilde{r}_\infty) &=& \left( -\frac{\tilde{p}_\infty \tilde{\Lambda}}{3} \tilde{r}_\infty^2 \right) e^{\tilde{Y}_\infty} \quad\; \approx \; 1 \,, \nonumber
	\\
	\tilde{\Phi}_\infty \; \approx \; 0 \quad &\text{and}& \quad \tilde{\Phi}'_\infty \; \approx \; 0 \,. \nonumber
\end{eqnarray}
\end{subequations}
	The vertical dashed gray line indicates the rescaled horizon radius $\tilde{r}_h$. The original horizon radius $r_h = 1$ is modified by the rescaling factor $e^{-\gamma \Phi_\infty / 2}$.
\begin{figure}[H]
	\centering
	\subfigure[][$\tilde{R}^2_{GB}$ vs. $\tilde{r}$]
{\includegraphics[width = 0.43 \textwidth]{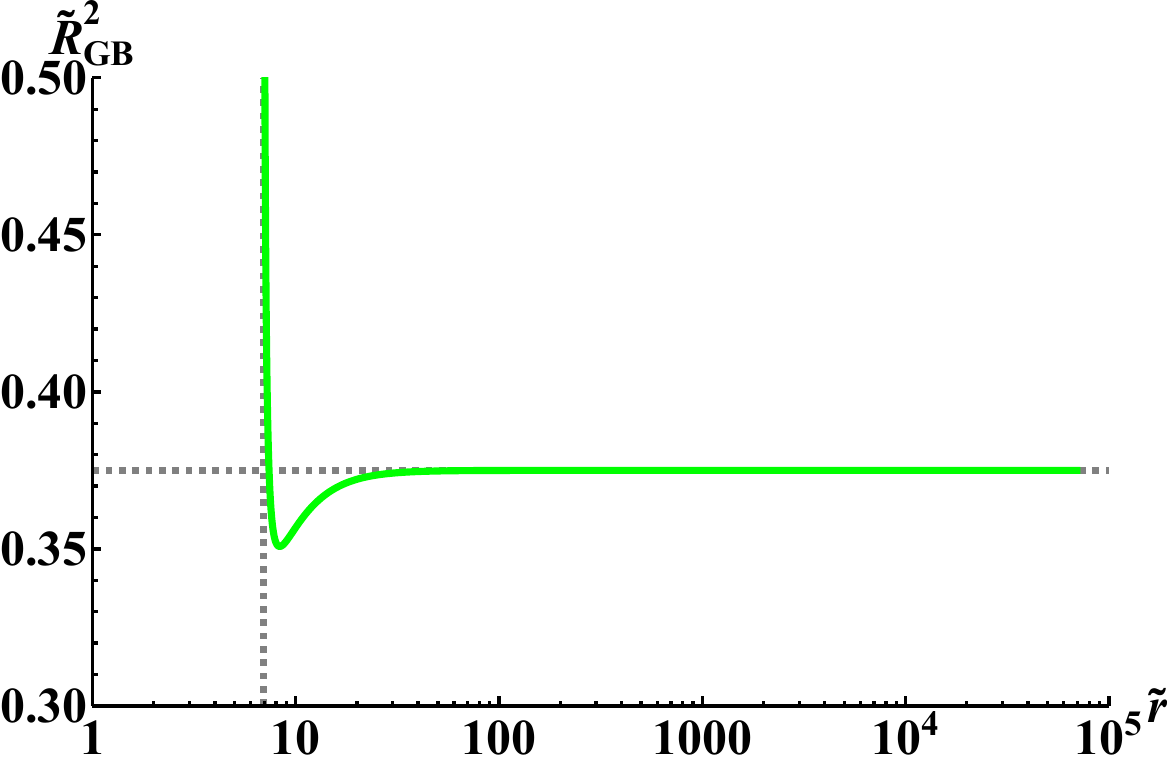}}
	\subfigure[][$2 \tilde{M}$ vs. $\tilde{r}$]
{\includegraphics[width = 0.43 \textwidth]{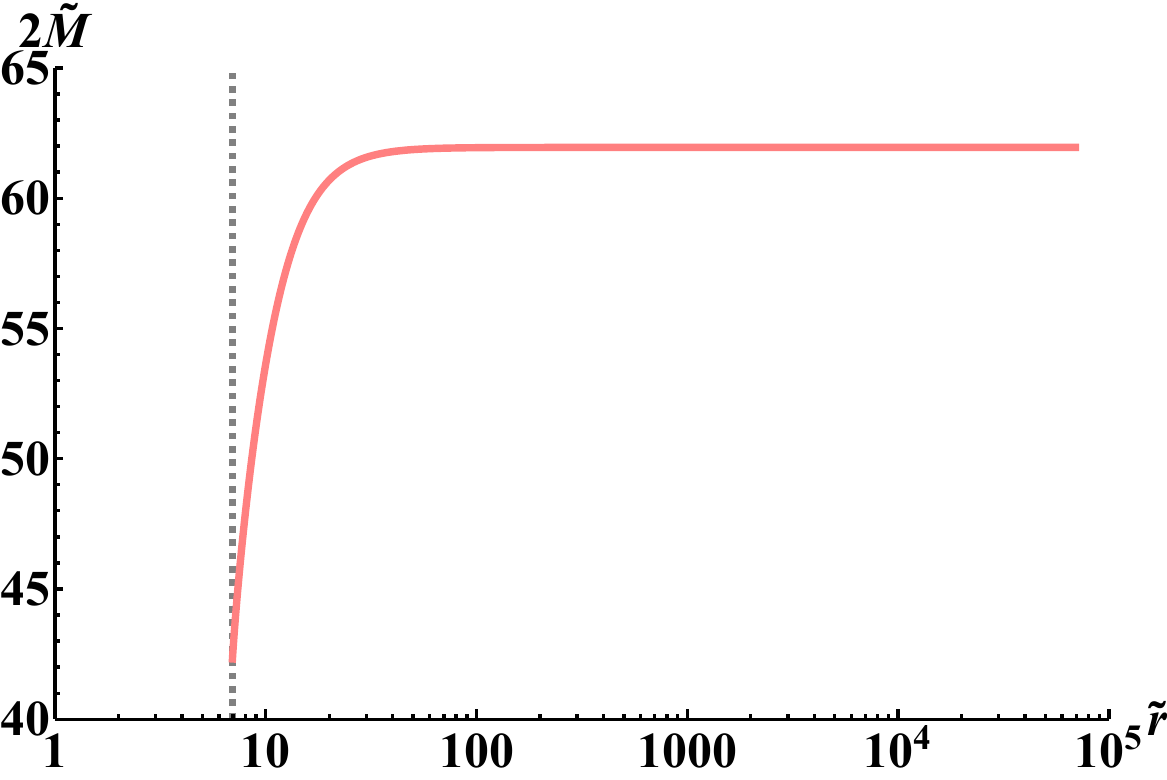}}
	\caption{The GB term $\tilde{R}^2_{GB}$ and mass function $2 \tilde{M}$.}
	\label{fig:RGB_2M}
\end{figure}
    Figure \ref{fig:RGB_2M} represents the GB term $\tilde{R}^2_{GB}$ and mass function $2 \tilde{M}$ with respect to $\tilde{r}$. In Fig. \ref{fig:RGB_2M}(a), the horizontal dashed gray line indicates $\frac{8}{3} \Lambda^2$ The GB term has a positive value in all regions of $\tilde{r}$ and converges to $\frac{8}{3} \Lambda^2$, which is calculated in Eq. \eqref{aRGB}. Only if the metric function conform to Schwarzschild-AdS-like solution in the asymptotic region, the mass function $2 \tilde{M}$ is derived from Eq. \eqref{aeY} as follows:
\begin{equation}
	2 \tilde{M} = \tilde{r} \left( 1 - \frac{\tilde{p} \tilde{\Lambda}}{3} \tilde{r}^2 - e^{-\tilde{Y}} \right)
	\label{2M}
\end{equation}
	and the ADM mass is represented as follows:
\begin{equation}
	M = M(r_h) + M_{hair}
	\label{ADM2M}
\end{equation}
	where the first term is the mass inside the horizon and the second term is the mass that comes from the dilaton field. In Fig. \ref{fig:RGB_2M}(b), the mass function also can be divided into two parts: the mass inside the horizon and the mass comes from the dilaton field. Since the dilaton field is decaying in the asymptotic region, the mass function increases and then converges to some constant value. The effect of the decaying dilaton field fades out as $\tilde{r}$ increases.
\begin{figure}[H]
	\centering
	\subfigure[][$\tilde{T}^t_{\ t}, \, \tilde{T}^r_{\ r}, \, \tilde{T}^\theta_{\ \theta} \enspace vs. \enspace \tilde{r}$]
{\includegraphics[width = 0.43 \textwidth]{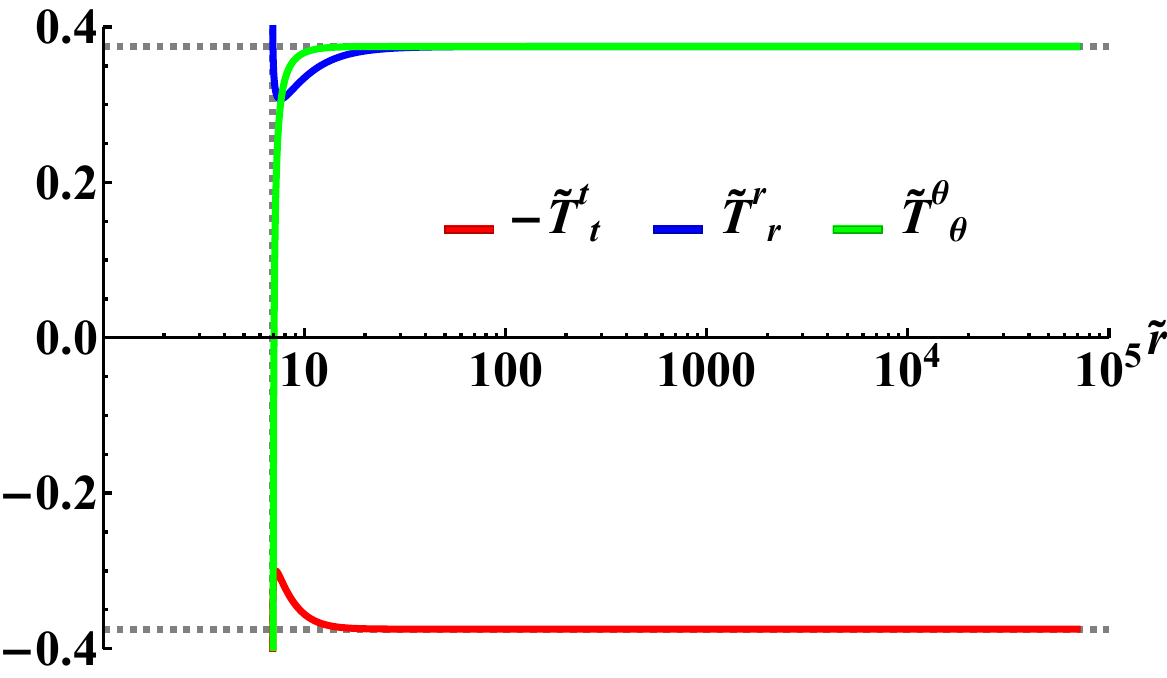}}
	\subfigure[][$-\tilde{T}^t_{\ t} + \tilde{T}^r_{\ r} \enspace vs.\enspace \tilde{r}$]
{\includegraphics[width = 0.43 \textwidth]{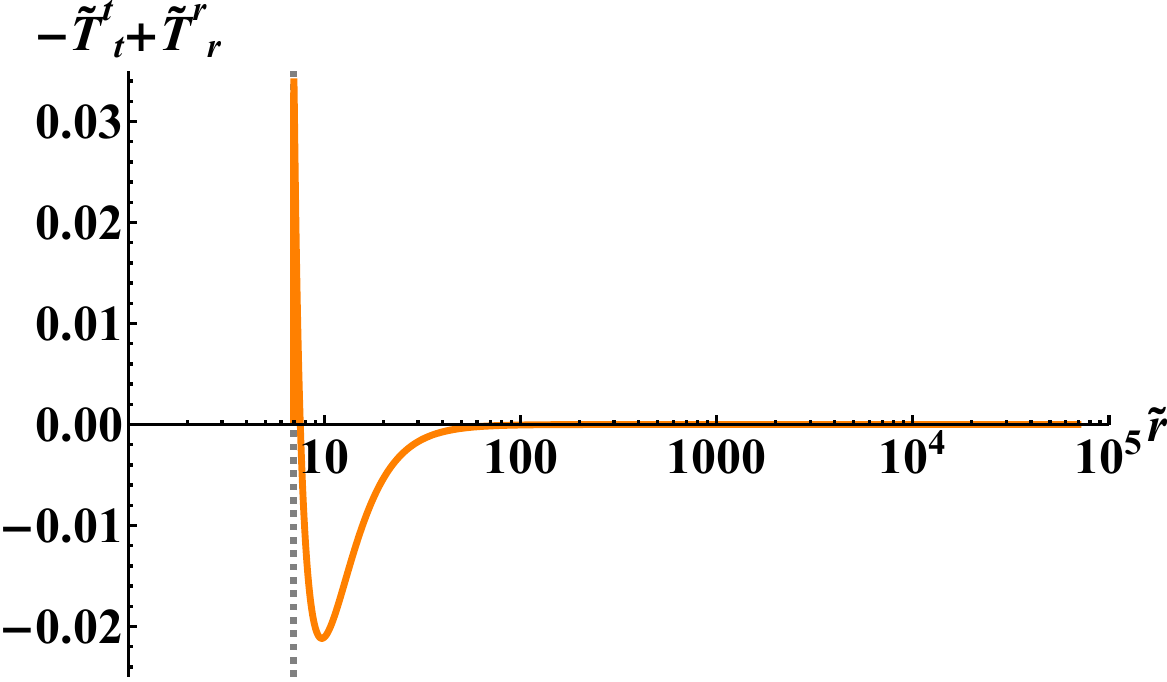}}
	\caption{The energy momentum tensor components and null energy condition.}
	\label{fig:EMT_Null}
\end{figure}
	Figure \ref{fig:EMT_Null} shows the components of the energy-momentum tensor and the value for the null energy condition. In Fig \ref{fig:EMT_Null}(a), the red line depicts the $(tt)$ component $-\tilde{T}^t_{\ t}$, the blue line depicts the $(rr)$ component $\tilde{T}^r_{\ r}$ and the green lines depict the ($\theta \theta$) component $\tilde{T}^\theta_{\ \theta}$. The $(tt)$ component and the $(rr)$ component start at the opposite value and converge to $\Lambda$ and $-\Lambda$, respectively, by Eq. \eqref{aTtt}, \eqref{aTrr} and \eqref{aTthth}. In Fig \ref{fig:EMT_Null}(b), the orange line depicts $-\tilde{T}^t_{\ t} + \tilde{T}^r_{\ r}$ to show the null energy condition. The negative part of $-\tilde{T}^t_{\ t} + \tilde{T}^r_{\ r}$ breaks the null energy condition.
	
	Taken together, the numerical solution with proper initial condition satisfies the near horizon behavior, which is characterized by divergent $X'$ and agrees with the Schwarzschild-AdS-like solution in the asymptotic region with decaying dilation field.

\newpage
\subsection{Black Hole in an AdS Spacetime with Varying $\gamma$ \label{sec3-3}}
	We obtain the solutions by the same numerical calculation by varying $\gamma$ where $\lambda + \gamma = 0$.
\begin{figure}[H]
	\centering
	\subfigure[][$-\left( k \tilde{r}^2 \right)^{-1} \tilde{g}_{tt}$s vs. $\tilde{r}$]
{\includegraphics[width = 0.43 \textwidth]{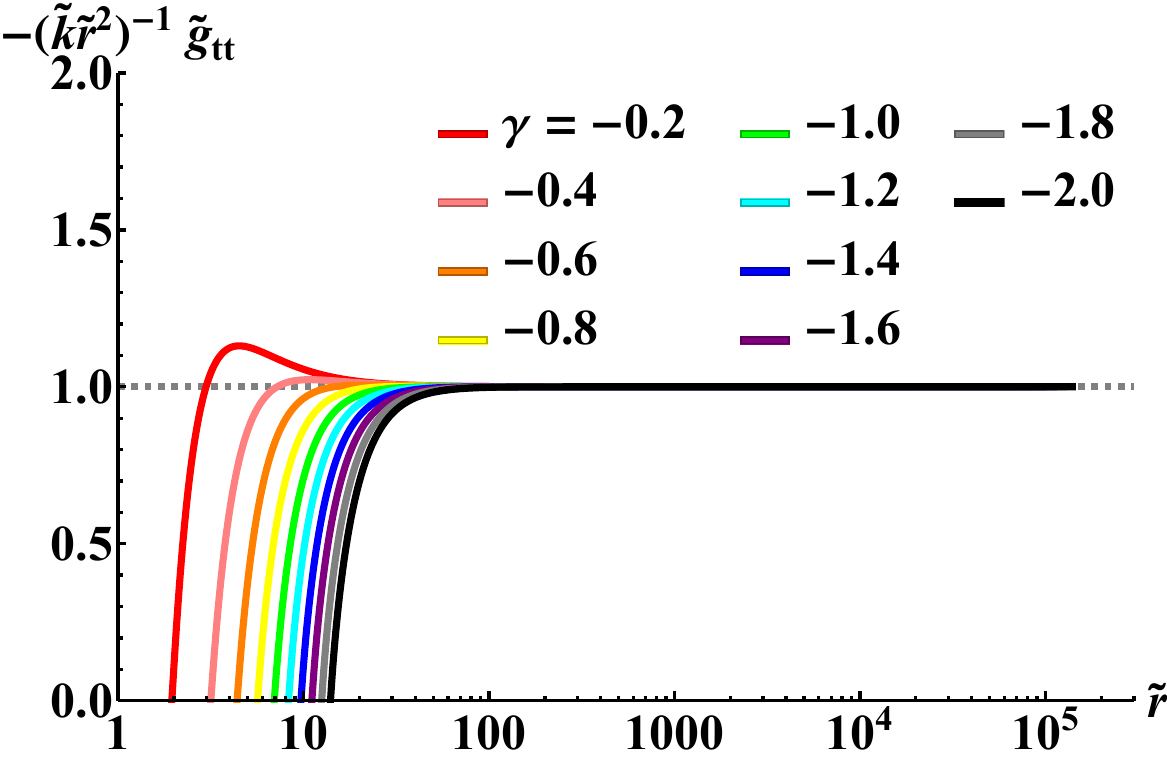}}
	\subfigure[][$k \tilde{r}^2 \tilde{g}_{rr}$s vs. $\tilde{r}$]
{\includegraphics[width = 0.43 \textwidth]{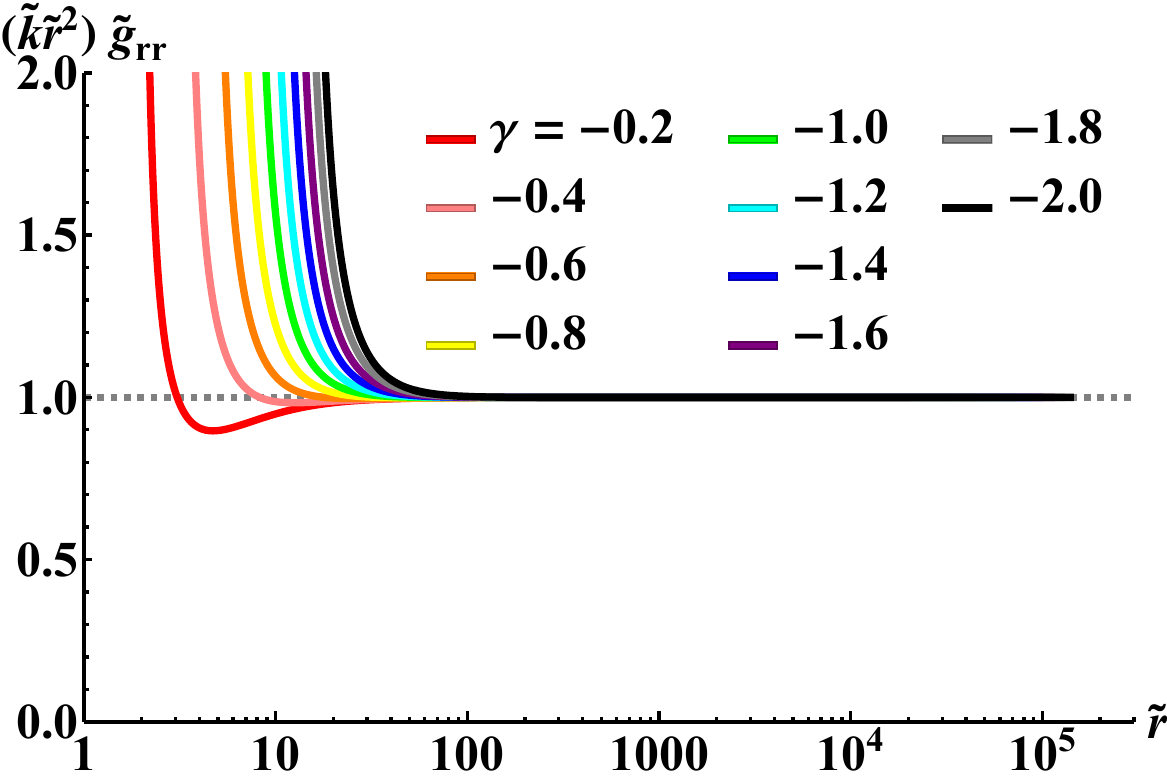}}
	\caption{The metric components with varying $\gamma$.}
	\label{fig:Metrics}
\end{figure}
	Figure \ref{fig:Metrics} illustrates the metric components with varying $\gamma$. All metric components fit well with the boundary conditions as Fig. \ref{fig:Metric_Phi}(a). We select $\gamma=-0.2$ for the red line,  $\gamma=-0.4$ for the pink line, $\gamma=-0.6$ for the orange line, $\gamma=-0.8$ for the yellow line,  $\gamma=-1.0$ for the green line, $\gamma=-1.2$ for the cyan line, $\gamma=-1.4$ for the blue line,  $\gamma=-1.6$ for the purple line, $\gamma=-1.8$ for the gray line, and $\gamma=-2.0$ for the black line. Each curve moving from the red to the black one corresponds to the case of a black hole whose mass increases with varying gamma and lambda.
\begin{figure}[H]
	\centering
	\subfigure[][$\tilde{\Phi}$s vs. $\tilde{r}$]
{\includegraphics[width = 0.43 \textwidth]{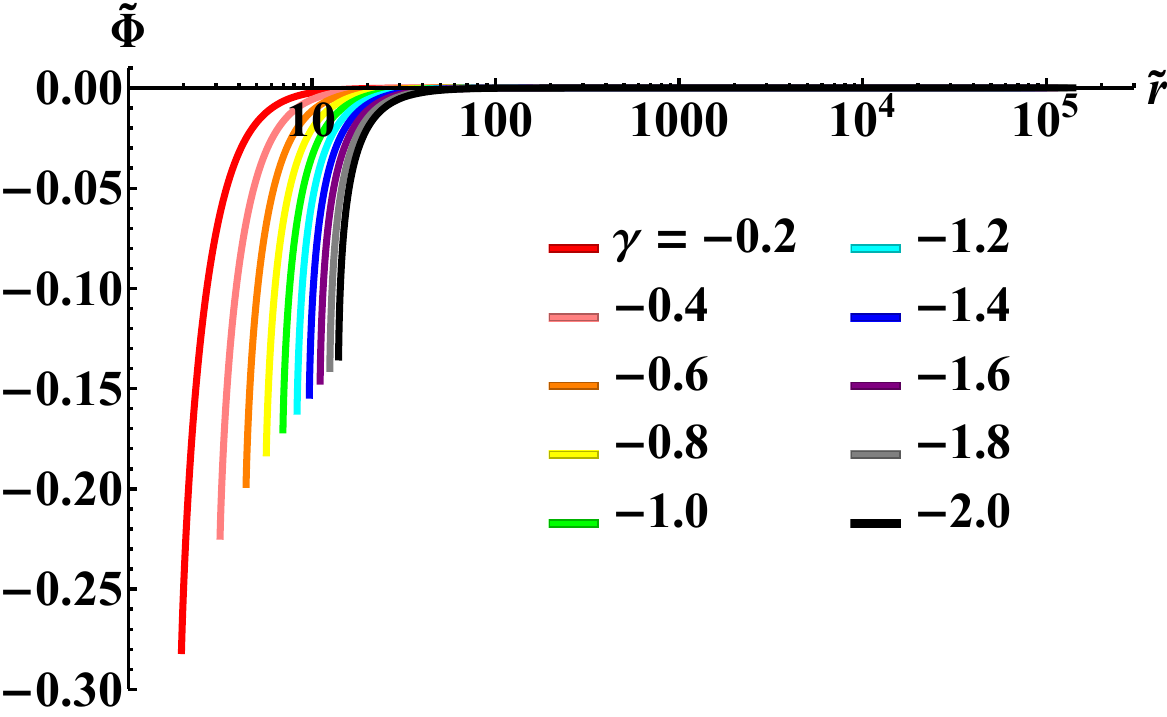}}
	\subfigure[][$\tilde{R}^2_{GB}$s vs. $\tilde{r}$]
{\includegraphics[width = 0.43 \textwidth]{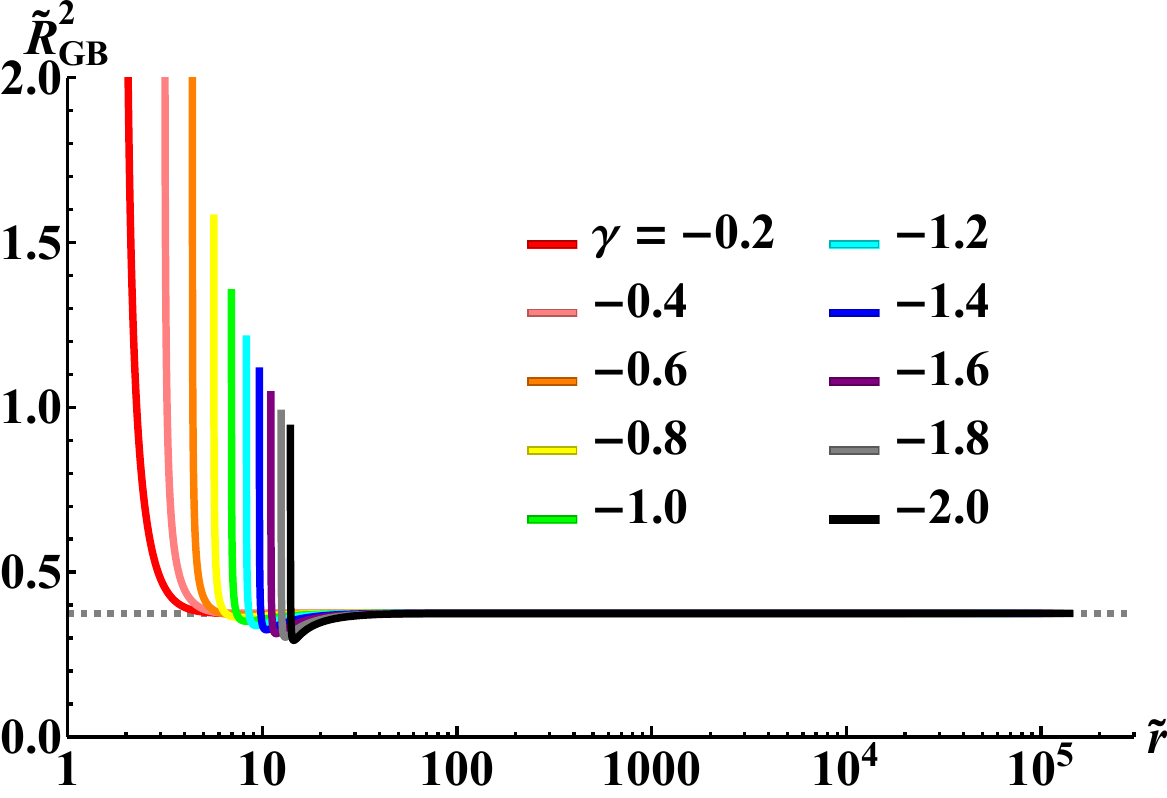}}
	\caption{The dilaton fields and GB terms with varying $\gamma$.}
	\label{fig:Phis_RGBs}
\end{figure}
    Figure \ref{fig:Phis_RGBs} represents the dilaton fields and GB terms with varying $\gamma$. In Fig. \ref{fig:Phis_RGBs}(a), every dilaton fields are decaying where $\tilde{r}$ increases. In Fig. \ref{fig:Phis_RGBs}(b), the GB terms are converged to the same constant value, $\frac{8}{3} \Lambda^2$, regardless of $\gamma$ by Eq. \eqref{aRGB}.
\begin{figure}[H]
	\centering
	\includegraphics[width = 0.5 \textwidth]{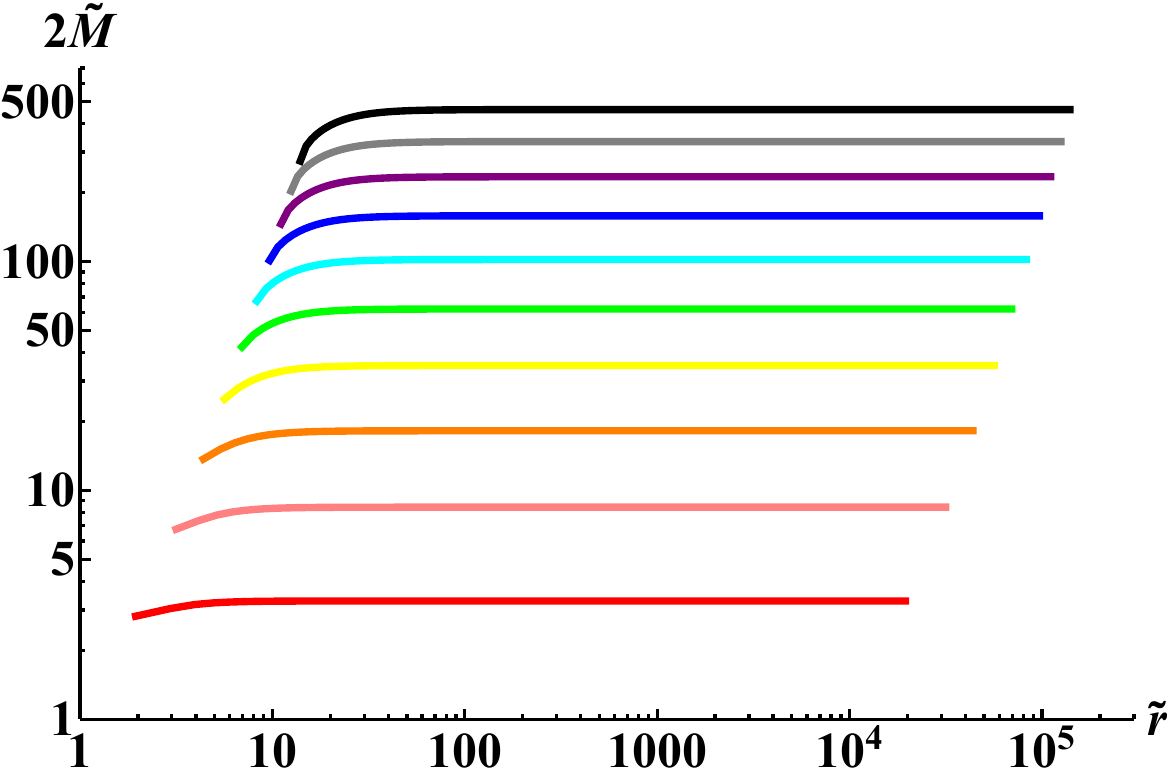}
	\caption{The mass functions with varying $\gamma$.}
	\label{fig:2Ms}
\end{figure}
    Figure \ref{fig:2Ms} shows the mass functions with varying $\gamma$. Similar to Fig. \ref{fig:RGB_2M}(b), each mass function converges to a constant value.

    Every solution follows the Schwarzschild-AdS-like solution in the asymptotic region under the proper condition that should be satisfied among the parameters with $\lambda + \gamma = 0$.

\section{Summary and Discussion \label{sec4}}
    We have constructed the hairy black hole solutions in DEGB theory in two cases: black hole solutions without the cosmological constant and with the negative $\alpha$, and with the negative cosmological constant and with the positive $\alpha$. We presented numerically obtained hairy black hole solutions with correction. We selected the suitable initial value for giving physical quantities without diverging at the event horizon. In \cite{Lee:2018zym}, the solution space was expand with the arbitrary sign of $\alpha$  by constructing the new integral constraint equation.

	In this article, we focused on constructing the AdS black hole solutions numerically and investigating them. We analyzed the components of the energy-momentum tensor. The energy density by the dilaton field has a negative value in the near horizon region, and this property does not satisfy one of the assumptions for the Bekenstein's novel no-hair theorem. We guess that this property, violation of the weak energy condition in a local region, seems to allow the scalar hair outside of the black hole horizon. We also checked the null-energy condition. The black hole solution in the asymptotically flat spacetime satisfies the null-energy condition in the entire spacetime, while the solution in AdS spacetime does not satisfy that. We guess that this property should be further studied concerning wormholes \cite{Kanti:2011yv, Bakopoulos:2020mmy}.

	In Sec. 2, we described the equations of motion and the procedure of solving them in detail. We expanded the metric function and the dilaton field in the near horizon limit and analyzed how the equations of motion with them need to be approximated in that limit. As a result, it came to know what value of the derivative of the dilation field should be chosen in that limit. We also analyzed what relation among the parameters should be satisfied when the dilaton field decays rapidly. We think this relation can be very useful in constructing black hole solutions in AdS spacetime.  Finally, the shifting and rescaling of the field were described in detail.

	In the last subsection, we obtained numerical solutions by varying the parameter $\gamma$ and showed the solution properties with varying that. The black hole mass of the corresponding numerical solution increases as the gamma value decreases. As the scalar hair decays, the value of the GB term approaches a constant one due to the influence of AdS spacetime.

	This article was prepared for the proceedings of the $17$th Italian-Korean Symposium for Relativistic Astrophysics.

\newpage
    \textit{Notes:} When this article was in the middle of publishing process, we recognized the paper \cite{Papageorgiou:2022umj}, in which the authors pointed out that there is an additional contribution coming from the boundary term, which is not vanish at the spatial infinity. We should have made the integral constraint with the vanishing boundary term. Accordingly, we present the modified integral constraint formula with correction and the vanishing boundary term, without changing any results, in the asymptotically flat spacetime:
    
    $0 = \int_{\mathcal M} d^4x \sqrt{-g} e^q \Phi \left[ \nabla^2 \Phi + \dot{q} R^2_{\rm GB} \right] = -\int_{\mathcal M} d^4x \sqrt{-g} \dot{q} e^q \Phi \left[ \left( 1 + \frac{1}{\dot{q} \Phi} \right) \left( \nabla \Phi \right)^2 - R^2_{\rm GB} \right] + \int_{\partial \mathcal M} d^3x \sqrt{-h} e^q \Phi n^\mu \nabla_\mu \Phi$.
    
    The rightmost term is the boundary term. Since, the dilaton field is static, only the radial part will contribute to the above equation. However, because of shifting, the dilaton field always vanishes at the spatial infinity. Hence, the boundary term also vanishes automatically.
    
\begin{figure}[H]
	\centering
	\includegraphics[width = 0.5 \textwidth]{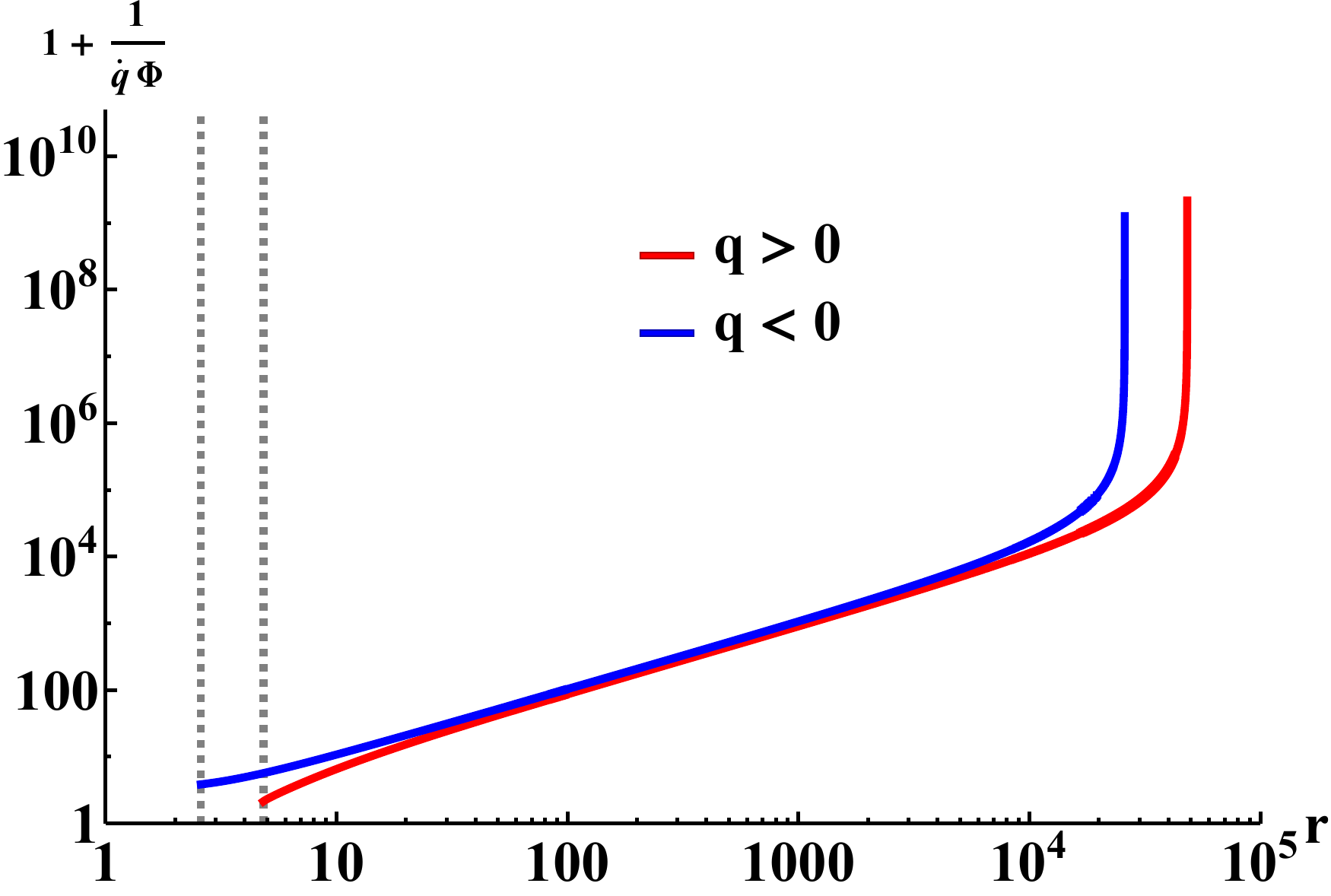}
	\caption{The factor attached to $(\nabla \Phi)^2$.}
	\label{fig:BD}
\end{figure}
    In the above equation, $\left( \nabla \Phi \right)^2$ and $R^2_{\rm GB}$ are both positive definite. Let us consider the following negative coupling function: $q(\Phi) = \alpha e^{\gamma \Phi}$, $\alpha < 0$ and $r_h^4 < 192 \alpha^2 \gamma^2$ for simplicity. At the horizon, $\dot{q}_h \Phi_h$ is positive as follows: $\dot{q}_h \Phi_h > \frac{\alpha}{2} e^{\gamma \Phi_h} \log\frac{r_h^4}{192 \alpha^2 \gamma^2} > 0$ and $\dot{q}_h \Phi_h$ is $0$ at the infinity. If we consider the dilaton field is monotonically increasing or decreasing with respect to radial length, $\dot{q} \Phi$ will monotonically decrease to $0$ and then $1 + \frac{1}{\dot{q} \Phi}$ will be positive in all regions. Figure 10 confirms numerically that the factor attached to $(\nabla \Phi)^2$, $1 + \frac{1}{\dot{q} \Phi}$, is always positive in both cases where $q$ is positive or negative. Therefore, the negative coupling function $q(\Phi)$ can satisfy the above equation and also can evade the no-hair theorem as a consequence.

\begin{acknowledgments}
	B.-H. L. (NRF-2020R1F1A1075472), W. L. (NRF-2022R1I1A1A01067336), and Center for Quantum Spacetime (CQUeST) of Sogang University (NRF-2020R1A6A1A03047877) were supported by Basic Science Research Program through the National Research Foundation of Korea funded by the Ministry of Education. We thank Seoktae Koh, Miok Park, and Gansukh Tumurtushaa for helpful comments on the appearance of the singular behavior in the solution and thank Miok Park again for discussion on the non-vanishing boundary term at the CQUeST 2022 Workshop in Yeosu, Korea, June 27\textemdash July 01, 2022. Thus, we fixed in this article. We are grateful to the organizers for their hospitality at the $17$th Italian-Korean Symposium on Relativistic Astrophysics in Kunsan National University, Korea, August 2\textemdash 6, 2021.
\end{acknowledgments}

\newpage
\bibliography{HBHDEGBBIB}

\end{document}